\documentclass[]{pasj01}
\begin{document} 
\Received{}
\Accepted{}

\title{Detached and Continuous Circumstellar Matter in Type Ibc Supernovae from Mass Eruption}

\author{Daichi \textsc{Tsuna}\altaffilmark{1,2}
\email{tsuna@caltech.edu}}
\altaffiltext{1}{TAPIR, Mailcode 350-17, California Institute of Technology, Pasadena, CA 91125, USA}
\altaffiltext{2}{Research Center for the Early Universe (RESCEU), School of Science, The University of Tokyo, 7-3-1 Hongo, Bunkyo-ku, Tokyo 113-0033, Japan}
\author{Yuki \textsc{Takei}\altaffilmark{2,3}
}
\altaffiltext{3}{Astrophysical Big Bang Laboratory, RIKEN, 2-1 Hirosawa, Wako, Saitama 351-0198, Japan}


\KeyWords{circumstellar matter --- supernovae: general --- stars: mass-loss} 

\maketitle

\begin{abstract}
Some hydrogen-poor supernovae (SNe) are found to undergo interaction with dense circumstellar matter (CSM) that may originate from mass eruption(s) just prior to core-collapse. We model the interaction between the remaining star and the bound part of the erupted CSM that eventually fall back to the star. We find that while fallback initially results in a continuous CSM down to the star, feedback processes from the star can push the CSM to large radii of $\gtrsim 10^{15}$ cm from several years after the eruption. In the latter case, a tenuous bubble surrounded by a dense and detached CSM extending to $\gtrsim 10^{16}$ cm is expected. Our model offers a natural unifying explanation for the diverse CSM structures seen in hydrogen-poor SNe, such as Type Ibn/Icn SNe that show CSM signatures soon after explosion, and the recently discovered Type Ic SNe 2021ocs and 2022xxf (the ``Bactrian") with CSM signatures seen only at late times.
\end{abstract}

\section{Introduction} 
Observations of supernovae (SNe) have revealed the intriguing end phases of massive stars, many with dense circumstellar matter (CSM) indicating mass loss much stronger than predicted from conventional theory (e.g., \cite{Kiewe12,Taddia13,Khazov16,Forster18,Morozova18,Bruch21}). The origin(s) of these dense CSM is unknown, but it may be related to the rapid increase of the nuclear energy generation rate in the final moments of a massive star's life (e.g., \cite{Quataert12,Smith14}).

Early observations have found signatures of dense CSM mainly in hydrogen-rich SNe through their narrow emission lines (Type IIn; \cite{Schlegel90}). Recent high-cadence surveys have found their existence in hydrogen-free SN Ibc, but with a diversity. Narrow lines of helium or carbon/oxygen are seen from days after explosion in a fraction of SN Ibc, which are respectively categorized as Type Ibn \citep{Pastorello08,Hosseinzadeh17} or Icn \citep{GalYam22,Pellegrino22,Perley22,Nagao23}. Intriguingly, a novel class of Type Ic SNe (``SN Ic-CSM") are discovered to show CSM signatures only after a month or two from explosion, which implies a CSM detached from the progenitor \citep{Ben-Ami14,Kuncarayakti22,Kuncarayakti23}. The diversity in the CSM structure of these Type Ibc SNe adds another mystery on the activity of massive stars during the final months to decades. 

 \begin{figure*}
 \centering
 \includegraphics[width=0.9\linewidth]{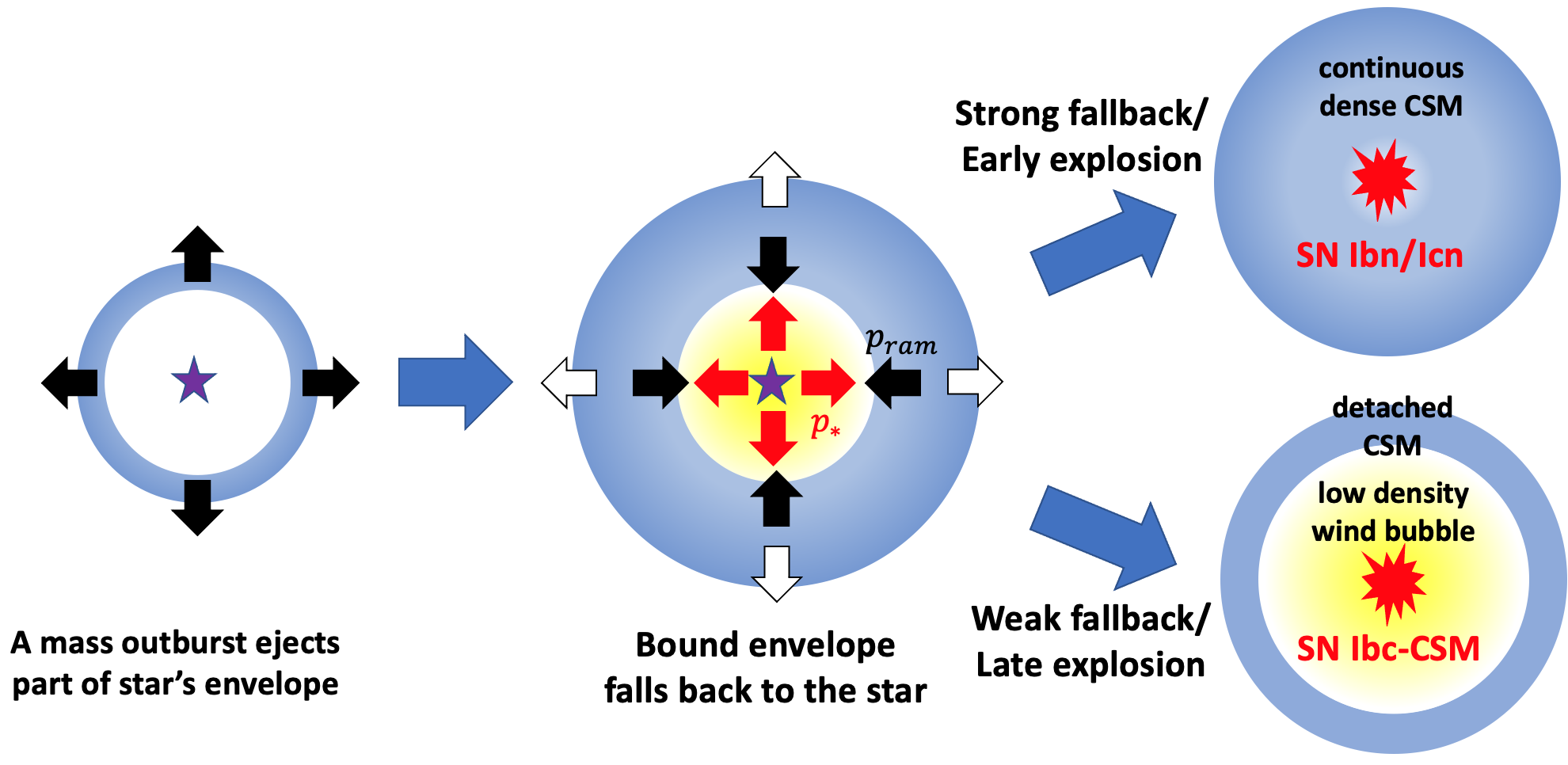}
\caption{Schematic picture of our model for the diverse structure of hydrogen-poor CSM in interacting Type Ibc SNe. Whether the CSM is detached at core-collapse is set by the competition between the ram pressure of the infalling CSM and the outward pressure from the remnant star.}
 \label{fig:schematic}
 \end{figure*}

In this work, we claim that this diversity can be naturally understood for CSM originating from eruptive mass loss. We show that dense CSM created by mass eruption would contain a bound part that falls back to the remnant star. We find that the outward pressure from the star can typically repel the fallback CSM from years after its eruption. This can naturally explain the observed diversity in the timing of the onset of CSM interaction in the SN phase.

This letter is constructed as follows. In Section \ref{sec:model} we describe our model in detail, and investigate how the properties of the CSM and the star affect the outcome of the CSM structure. In Section \ref{sec:Discussion} we discuss this model in the context of observed SNe with hydrogen-poor CSM, and mention possibilities to test and refine of our model.

\section{Our model}
\label{sec:model}

 We show a schematic picture of our model in Figure \ref{fig:schematic}. We assume that the dense CSM is created by eruptive mass-loss \citep{Dessart10,Owocki19,Kuriyama20,Linial21,Ko22,Tsang22} of a stripped star that ejects part of its envelope. While we remain agnostic to its mechanism in this work, pre-SN outbursts of hydrogen-poor stars may occur due to e.g. wave-heating \citep{Fuller18,Leung21} or pulsational pair-instability \citep{Yoshida16,Leung19,Renzo20}. 
 
 For partial explosions, an inner bound part would generally exist and fall back towards the leftover star. We first model mass eruption of stripped stars using radiation hydrodynamical simulations and show that fallback of the bound CSM robustly occur. We then analytically model the interplay between the star and the fallback CSM, and discuss the expected density profile at core-collapse.
 
\subsection{Mass Eruption from Progenitors of Type Ibc SNe}
 \begin{figure*}
 \centering
 \includegraphics[width=0.9\linewidth]{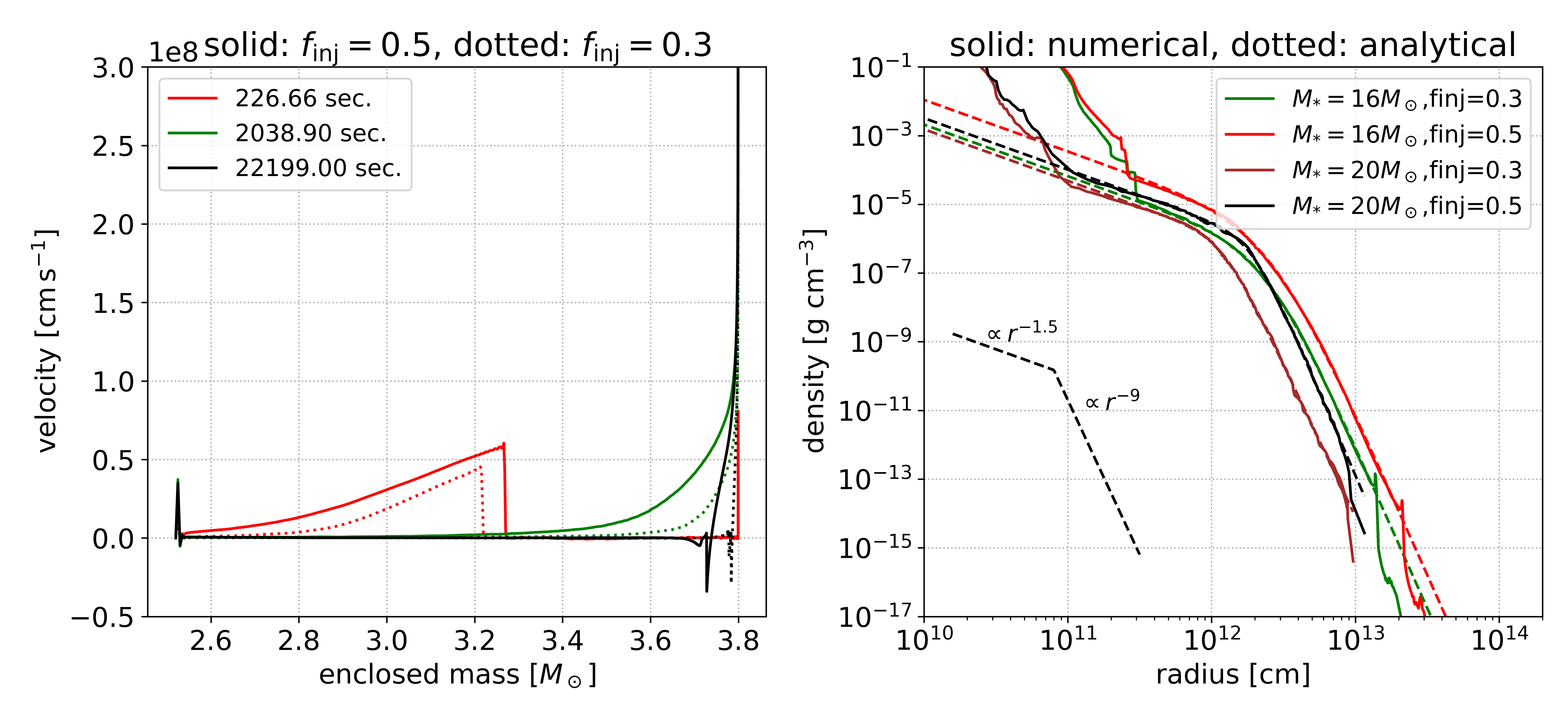}
\caption{Summary of the mass eruption done by CHIPS. (Left panel) Velocity as a function of enclosed mass for the helium star model at 3 epochs, with solid (dashed) lines for $f_{\rm inj}=0.5\ (0.3)$. (Right panel) CSM profiles at the end of the simulation. Dashed lines show the non-linear least square fits by equation \ref{eq:rho_CSM}.}
 \label{fig:Mr_vs_v_and_r}
 \end{figure*}

As examples of the progenitors for interacting SN Ibc, we prepare hydrogen-poor and helium-poor pre-SN models with zero-age main sequence masses of $16,\,20\ M_\odot$ respectively using \verb|MESA| version 12778 \citep{Paxton11,Paxton13,Paxton15,Paxton18,Paxton19}. The model parameters basically follow the test suite \verb|example_make_pre_ccsn|. For the $16\ M_\odot$ model we strip the hydrogen envelope at the end of core hydrogen burning, while for the $20\ M_\odot$ model we strip both hydrogen and helium layers after core helium burning. The resulting masses $M_*$, radii $R_*$ and luminosities $L_*$ of the two stars are ($3.8M_\odot$, $3.0R_\odot$, $3.9\times 10^4L_\odot$) and ($4.7M_\odot$, $0.22R_\odot$, $1.3\times 10^5L_\odot$) respectively.

We model mass eruption using the open-source code CHIPS (\cite{Kuriyama20,Takei22}, Takei et al. 2023 in prep.), which solves eruption of the envelope using 1D Lagrangian radiation hydrodynamical simulations. The envelope is partially erupted by sudden energy injection at its base, with energy scaled by the envelope's binding energy of $f_{\rm inj}$. For the helium-rich star we extract the helium layer, while for the helium-poor star we extract the CO layer just above the silicon core. We choose $f_\mathrm{inj}=0.3$ and $0.5$, which result in masses of the CSM of order $0.01$--$0.1\ M_\odot$ in line with those observed in interacting Type Ibc SNe (see also Figure \ref{fig:profile_noFB}). We stop the simulations at $2.2\times 10^4$ s for the $16M_\odot$ and $3.8\times 10^3$ s for the $20M_\odot$ model, when the bulk of the CSM reaches $\approx 10^{12}$ cm and is expanding nearly homologously.

We plot the resulting shock propagation and CSM density profile in Figure \ref{fig:Mr_vs_v_and_r}. The eruption occurs $10^2$--$10^3$s after energy injection, and a fraction of the envelope is ejected. At times much later than the star's free-fall time
\begin{eqnarray}
    t_{\rm ff,*} \approx \left(\frac{R_*^3}{GM_*}\right)^{1/2} \sim 4500\ {\rm s}\left(\frac{R_*}{3R_\odot}\right)^{3/2} \left(\frac{M_*}{3M_\odot}\right)^{-1/2},
\end{eqnarray}
the internal pressure in the CSM has negligible influence on its motion. Neglecting other sources of pressure, the CSM density profile at this regime simply reflects the trajectory of the CSM under the gravitational pull from the central star. Subsequently the inner bound part of the CSM falls back, as seen in the left panel of Figure \ref{fig:Mr_vs_v_and_r}. The resulting density profile was analytically and numerically found by \citet{Tsuna21} to follow a double power-law
\begin{eqnarray}
\rho_{\rm CSM}(r) = \hat{\rho}_{\rm CSM} \left[\frac{(r/r_{\rm CSM})^{1.5/y}+(r/r_{\rm CSM})^{n_{\rm out}/y}}{2}\right]^{(-y)},
\label{eq:rho_CSM}
\end{eqnarray}
where $r_{\rm CSM}, \hat{\rho}_{\rm CSM}$ are radius and density at the transition, and $n_{\rm out}$ is the power-law index of the outer part of the CSM, and $y$ is a constant. This profile approaches a power-law $\rho\propto r^{-1.5}$ at $r\ll r_{\rm CSM}$ characterized by the fallback, while the outer homologous part follows a steep power-law $\rho\propto r^{-n_{\rm out}}$ like the ejecta from SNe. While this was only verified for hydrogen-rich progenitors in \citet{Tsuna21}, Figure \ref{fig:Mr_vs_v_and_r} shows that it can also fit well the profile of CSM from stripped stars, with $n_{\rm out}\approx 9$--$10$ and $y\approx 2$--$4$.

\subsection{Fallback CSM versus Radiation Pressure}
\label{sec:Condition}
The profile above results from the trajectory of the CSM simply under the influence of gravity. However, radiation pressure from the central star can influence the fallback of the bound CSM at sufficiently late times, which can significantly modify the CSM profile. 

Adopting $n_{\rm out}=9$ and $y=3$, the density $\hat{\rho}_{\rm CSM}$ at $r_{\rm CSM}$ is related to the total CSM mass $M_{\rm CSM}$, obtained by integrating mass from $r=0$ to infinity, as
\begin{eqnarray}
    \hat{\rho}_{\rm CSM} &\approx& {M_{\rm CSM}}/{37r_{\rm CSM}^3} \nonumber \\
            &\sim& 1.7\times 10^{-16}{\rm g\ cm^{-3}} \nonumber \\
            &&\times  \left(\frac{M_{\rm CSM}}{0.1\ M_\odot}\right) \left(\frac{v_{\rm CSM}}{10^3\ {\rm km\ s^{-1}}}\right)^{-3}\left(\frac{t}{1\ {\rm yr}}\right)^{-3},
\end{eqnarray}
where $v_{\rm CSM}=r_{\rm CSM}/t$ is the velocity of the CSM at $r=r_{\rm CSM}$ that expands homologously. We hereafter fix this velocity as $v_{\rm CSM}=10^3\ {\rm km\ s^{-1}}$, a value commonly adopted for the CSM of stripped SNe.

The fallback CSM lies within the radius $r_{\rm ff}$ where the free-fall timescale is equivalent to $t$
\begin{eqnarray}
    r_{\rm ff} \approx (GM_*t^2)^{1/3} \sim 7.3\times 10^{13}\ {\rm cm} \left(\frac{M_*}{3M_\odot}\right)^{1/3}\left(\frac{t}{1\ {\rm yr}}\right)^{2/3}.
\end{eqnarray}
which is much less than $r_{\rm CSM}$ for $t\gg t_{\rm ff,*}$. The density and velocity profile in the fallback region ($r\lesssim r_{\rm ff}$) thus follows 
\begin{eqnarray}
    \rho_{\rm fb}(r<r_{\rm ff})&\approx& \hat{\rho}_{\rm CSM}(r/r_{\rm CSM})^{-1.5} \propto r^{-1.5}t^{-1.5} \\
    v_{\rm fb}(r<r_{\rm ff}) &\approx& -\sqrt{2GM_*/r} \propto r^{-0.5}.
\end{eqnarray}
The luminosity of the remnant star after eruption is uncertain, because the star's thermal adjustment timescale
\begin{eqnarray}
    \frac{GM_*^2}{R_*L} \sim 1000\ {\rm yr}\left(\frac{M_*}{3M_\odot}\right)^2\left(\frac{R_*}{3R_\odot}\right)^{-1} \left(\frac{L}{10^5L_\odot}\right)^{-1}
\end{eqnarray}
is generally much longer than the time left until core-collapse, which is decades or less for these CSM. We formulate the luminosity as $L_*=\Gamma L_{\rm Edd}$, where 
\begin{eqnarray}
    L_{\rm Edd}&=&4\pi GM_*c/\kappa \nonumber \\
    &\approx& 7.5\times 10^{38}\ {\rm erg\ s^{-1}}\left(\frac{M_*}{3M_\odot}\right)\left(\frac{\kappa}{0.2\ {\rm cm^2\ g^{-^1}}}\right)^{-1}
\end{eqnarray}
is the Eddington luminosity and $\Gamma$ is the Eddington ratio. We assume a fixed opacity of $\kappa = 0.2\ {\rm cm^2\ g^{-^1}}$ corresponding to ionized hydrogen-free gas.

For a shell of material at radius $r_{\rm sh}$, velocity $v_{\rm sh}$ and mass $M_{\rm sh}$ in the fallback region, the equation of motion is given as (e.g. Section 4.2 of \cite{Sakurai16})
\begin{eqnarray}
    \frac{d}{dt}(M_{\rm sh}v_{\rm sh}) &=& P_* - \frac{GM_{\rm sh}M_*}{r_{\rm sh}^2} - \dot{M}_{\rm fb}(v_{\rm sh}-v_{\rm fb}) \label{eq:dMvdt}\\
    \frac{d}{dt}(M_{\rm sh}) &=& \dot{M}_{\rm fb}=4\pi r_{\rm sh}^2 \rho_{\rm fb} (v_{\rm sh}-v_{\rm fb}) \label{eq:dMdt}
\end{eqnarray}
where the first term in the momentum equation is from the star, second term is the gravity, and the last term is the ram pressure from the CSM. For the first term we consider contributions from radiation pressure and stellar wind,
\begin{eqnarray}
    P_* = {\rm min}(\tau_{\rm sh}, 1)\frac{L_*}{c} + \dot{M}_wv_w \equiv \left[{\rm min}(\tau_{\rm sh}, 1) + \eta_w\right]\frac{\Gamma L_{\rm Edd}}{c} 
\end{eqnarray}
where $\tau_{\rm sh}\approx \kappa M_{\rm sh}/4\pi r_{\rm sh}^2$ is the shell's optical depth, and $\dot{M}_w$ and $v_w$ are respectively the mass-loss rate and velocity of the stellar wind. We have parameterized the contribution from the wind by a single ``wind efficiency parameter" $\eta_w\equiv \dot{M}_wv_w/(L_*/c)$, which is larger for higher $\Gamma\propto L_*/M_*$ (e.g., \cite{Vink00,Sander20}) \footnote{To reduce the parameter space we neglected the contribution from the wind $\dot{M}_w$ in equation (\ref{eq:dMdt}). However including this term, assuming a typical value of $\dot{M}_w=10^{-5}\ M_\odot\ {\rm yr}^{-1}$, did not affect the results when modelling the shell evolution in Section \ref{sec:evolution}.}.

Considering the shell is at rest ($v_{\rm sh}=0$) at $r_{\rm sh}=R_*$, the ram pressure term can be evaluated as
\begin{eqnarray}
    4\pi r^2 \rho_{\rm fb} v_{\rm fb}^2 \approx \frac{8\pi GM}{R_*^{1/2}} \hat{\rho}_{\rm CSM} r_{\rm CSM}^{1.5} \approx \frac{2\kappa M_{\rm CSM}}{37R_*^{1/2}v_{\rm CSM}^{3/2}t^{3/2}}\frac{L_{\rm Edd}}{c}
\end{eqnarray}
which falls with time as $\propto t^{-1.5}$. The outward pressure from the star has to be roughly twice as big as the ram pressure in order to overcome the gravitational pull. This is achieved from a time after eruption of
\begin{eqnarray}
    t_{\rm crit} &\approx& 8.9\ {\rm years}\ \left[{\rm min}(\tau_{\rm sh}, 1)\Gamma + \eta_w\Gamma\right]^{-2/3} \nonumber \\
    &&\times \left(\frac{R_*}{3R_\odot}\right)^{-1/3} \left(\frac{M_{\rm CSM}}{0.1\ M_\odot}\right)^{2/3}\left(\frac{v_{\rm CSM}}{10^3\ {\rm km\ s^{-1}}}\right)^{-1}.
\end{eqnarray}
The scaling indicates that typically the CSM has to expand for years until it can be detached from the star, and this wait time would be longer for larger $M_{\rm CSM}$ and/or dimmer progenitors with low $\Gamma$ (and $\eta_w$). The ram pressure falls with radius as $\propto r^{-1/2}$, so once $P_*$ can overcome the ram pressure at $r=R_*$ it can do so at larger radii as well.

The dependence of the wind properties on $\Gamma$ has recently been investigated for stripped stars in the helium main sequence phase \citep{Vink17,Sander20}. The value of $\eta$ can exceed unity for stars with $\Gamma \gtrsim 0.2$ at solar metallicity, and can reach $\approx 10$ for $\Gamma=0.5$ (Figure 3 of \cite{Sander20}). We note that mass-loss can be different for these stars close to core-collapse, and may be even stronger (e.g., \cite{Moriya22}). This nevertheless implies that luminous/massive progenitors are more likely to detach the CSM at earlier times.

\subsection{CSM Structure at Core-collapse}
\label{sec:evolution}
Having seen the condition for the CSM to be pushed back by the star, we consider the CSM structure at core-collapse for the cases when the CSM cannot and can be repelled.

\begin{figure}
 \centering
  \includegraphics[width=\linewidth]{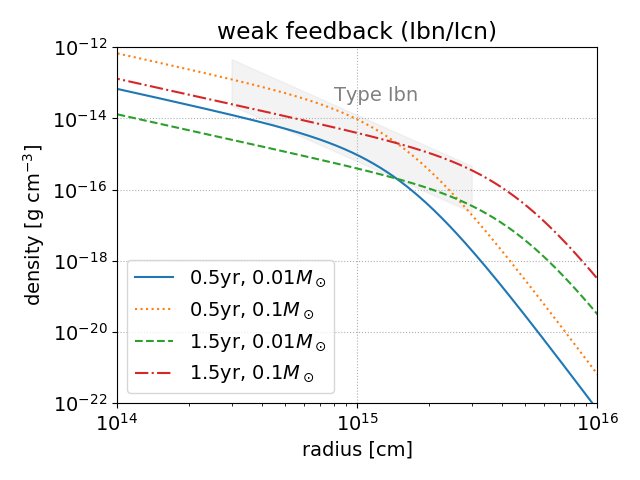}
\caption{CSM Density profile in the case of weak stellar feedback (equation \ref{eq:rho_CSM} assuming $n_{\rm out}=9$ and $y=3$), with varied total CSM mass and time from eruption to core-collapse. The gray shaded region indicates the density profile inferred from optical light curves of SN Ibn \citep{Maeda22}.}
 \label{fig:profile_noFB}
 \end{figure}

For the former, we expect the profile to be largely characterized by the CSM moving under the star's gravity, as in equation (\ref{eq:rho_CSM}). In Figure \ref{fig:profile_noFB} we plot the density profile in equation (\ref{eq:rho_CSM}) for a four sets of $t$ and $M_{\rm CSM}$, assuming $v_{\rm CSM}=10^3$ km s$^{-1}$. The profile agrees with that inferred from light curve modelling of SN Ibn \citep{Maeda22}, for $t\approx 1$ year and $M_{\rm CSM}=0.01-0.1\ M_\odot$. We expect the profile to flatten around $10^{15}$ cm, which is seen in three out of seven SN samples in \citet{Maeda22}.

\begin{figure}
 \centering
  \includegraphics[width=\linewidth]{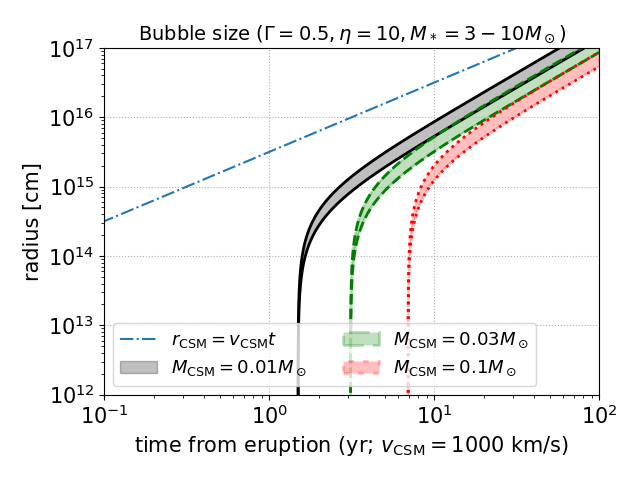}
\caption{Time evolution of the shell pushed by stellar feedback, which corresponds to the bubble radius. Shaded regions show the range for varied $M_*$ from $3M_\odot$ to $10M_\odot$. The dot-dashed line shows the characteristic CSM radii $r_{\rm CSM}$ for $v_{\rm CSM}=10^3\ {\rm km\ s^{-1}}$.}
 \label{fig:wind_vs_fallback}
 \end{figure}
 
For the case where the CSM is repelled, we estimate how far the CSM is detached by calculating the time evolution of the shell radius. We integrate equations (\ref{eq:dMvdt}), (\ref{eq:dMdt}), and $dr_{\rm sh}/dt = v_{\rm sh}$. The obtained $r_{\rm sh}(t)$ is the radius at time $t$ within which the fallback CSM is evacuated, and corresponds to the extent of the wind bubble in Figure \ref{fig:schematic}. The initial conditions at $t=t_{\rm crit}$ are set as $r_{\rm sh}=R_*$, $v_{\rm sh}=0$, and $M_{\rm sh}=0$. Any fallback matter before $t_{\rm crit}$ is assumed to be part of the star, and does not contribute to $M_{\rm sh}$.

Figure \ref{fig:wind_vs_fallback} shows the evolution for $(\Gamma, \eta) = (0.5, 10)$ and three CSM masses of $0.01M_\odot$, $0.03M_\odot$, and $0.1M_\odot$. The adopted $\Gamma$ roughly corresponds to that of the helium-poor star we adopted in Section \ref{sec:model}. The shaded regions show a range of $M_*$ from $3M_\odot$ to $10M_\odot$, with higher $M_*$ resulting in larger outward pressure and hence larger radii. We find that the CSM can be evacuated to a large radii of $>10^{15}$ cm  within 10 years from eruption, with a larger radii for lower CSM mass. The obtained radii at $t\approx 10$ yr are roughly consistent with estimates of the inner edge of the detached CSM, $\approx 4$--$5\times 10^{15}$ cm, for Type Ic SN 2021ocs and 2022xxf \citep{Kuncarayakti22,Kuncarayakti23}. 

\section{Discussion and Conclusion}
\label{sec:Discussion}
In this work we have shown that the diversity in the CSM structure of interacting Type Ibc SNe can be explained by the interplay between the fallback CSM and the feedback from the star. Such fallback is a natural consequence of outbursts that eject only a fraction of the envelope as CSM, which is expected for most interacting SNe.

If our interpretation is correct, the observed SN Ic-CSM with detached CSM imply (i) a longer interval from eruption to core-collapse of $\gtrsim$ years, and/or (ii) stronger stellar feedback compared to SN Ibn/Icn. The former is consistent with observations, since the duration of the interaction ($\gtrsim 100$ days) is longer compared to Ibn/Icn (10s of days). The latter may also agree with a small ejecta mass generally inferred for SN Ibn/Icn ($\lesssim$ a few $M_\odot$), favoring a light progenitor with small $\Gamma$ and $\eta$. The progenitors for SN Ic-CSM is not well constrained, but nebular spectra implies that they may be more massive than those in SN Ibn/Icn \citep{Kuncarayakti23}.

We conclude by discussing the future avenues to test and refine this model. The outburst that created the dense CSM would itself accompany some emission \citep{Dessart10,Kuriyama20,Matsumoto22,Tsuna23}, which would be crucial to distinguish this model from other mechanisms. Considering the plateau emission from cooling of the expanding CSM as it recombines \citep{Popov93, Kasen09}, we obtain a SN precursor luminosity
\begin{eqnarray}
    L_{\rm pre} &\sim& 4.5\times 10^{39}\ {\rm erg\ s^{-1}} \nonumber \\
    &&\times \left(\frac{E_{\rm CSM}}{10^{48}\ {\rm erg}}\right)^{5/6}\left(\frac{M_{\rm CSM}}{0.1\ M_\odot}\right)^{-1/2} \left(\frac{R_*}{5\ R_\odot}\right)^{2/3}
\end{eqnarray}
where $E_{\rm CSM}\approx M_{\rm CSM}v_{\rm CSM}^2/2$ is the kinetic energy of the unbound CSM, and we assumed a helium-rich gas with opacity $\kappa=0.2\ {\rm cm^2\ g^{-1}}$ and recombination temperature $10^4$ K following \citet{Fernandez18}. Such low $L_{\rm pre}$ is consistent with non-detections for most Type Ibn/Icn SNe. An exception is a Type Ibn SN 2006jc, where a luminous precursor of $\approx -14$ mag ($L_{\rm pre}\sim 10^{41}$ erg) was detected 2 years before core-collapse \citep{Nakano06,Pastorello07}. Stripped stars in a narrow mass range around $3\ M_\odot$ are found to expand to as large as $\sim 100\ R_\odot$ \citep{Woosley19,Ertl20,Wu22}, which may explain the large $L_{\rm pre}$ for this event. SN 2006jc-like outbursts, and possibly dimmer ones, are within reach for optical surveys like the Zwicky Transient Facility \citep{Bellm19} or the Rubin observatory \citep{Ivezic19}.

In this work we have also not made predictions on the emission from the interacting SN following core-collapse. Detailed light curve modelling with both CSM interaction and radioactive decay of $^{56}$Ni would be important to accurately diagnose the structure of the CSM, and will be reported in a forthcoming study (Takei 2023 et al. in prep.).

Our analytical estimation for the CSM contains approximations, and detailed numerical simulations is desired to more accurately predict the density profile. A major uncertainty of our model is the influence of fallback CSM on the launch and acceleration of the stellar wind. Furthermore, multi-dimensional effects such as Rayleigh-Taylor instabilities can become important when modelling the interaction between the fallback and wind/radiation.

\begin{ack}
DT thanks Jim Fuller, Toshikazu Shigeyama and Akihiro Suzuki for discussions. DT is supported by the Sherman Fairchild Postdoctoral Fellowship at Caltech. 
\end{ack}

\bibliographystyle{apj} 
\bibliography{CSM}

\end{document}